\documentclass[12pt]{article}
\usepackage[utf8]{inputenc}%agregado
\usepackage{amsmath}
\usepackage{graphicx}
\usepackage{amssymb}%agregado
\usepackage{color}	
\usepackage{soul}%para tachar

\usepackage{float}%poner caption arriba
\floatstyle{plaintop}
\restylefloat{table}

\usepackage{multirow}%para unir filas

\usepackage[square,numbers]{natbib}%agregado ref apa

\usepackage[section,nottoc]{tocbibind}%poner referenica en la tabla de contenidos

\usepackage[breaklinks=true,backref=page]{hyperref}%poner hiperreferencias en el pdf

\title{Simple models for macro-parasite distributions in hosts}
\author{Gonzalo Maximiliano LOPEZ$^{1,3,4}$, Juan Pablo APARICIO$^{1,2}$\\
\\
{\small $^1$ Instituto de Investigaciones en Energ\'ia no Convencional (INENCO),} \\ {\small Consejo Nacional de Investigaciones Cient\'ificas y T\'ecnicas (CONICET),}\\{\small Universidad Nacional de Salta, Av. Bolivia 5150, 4400 Salta, Argentina.}\\
{\small $^2$ Simon A. Levin Mathematical, Computational and Modeling Sciences Center,} \\ {\small Arizona State University, PO Box 871904 Tempe, AZ 85287-1904, USA}\\
{\small $^3$ Departamento de Matem\'atica,}\\{\small Universidad Nacional de Salta, Av. Bolivia 5150, 4400 Salta, Argentina.}\\
{\small $^4$ Corresponding author: gonzalo.maximiliano.lopez@gmail.com}}
\date{}

\newcommand{\minitab}[2][l]{\begin{tabular}{#1}#2\end{tabular}}%para la tabla

\begin{document}
\maketitle
%\newpage
%\tableofcontents

\begin{abstract}
\addcontentsline{toc}{section}{Abstract}
Negative binomial distribution is the most used distribution to model macro-parasite burden in hosts. However reliable maximum likelihood parameter estimation from data is far from trivial. No closed formula is available and numerical estimation requires sophisticated methods. Using data from the literature we show that simple alternatives to negative binomial, like zero-inflated geometric or hurdle geometric distributions, produce a good and even better fit to data than negative binomial distribution. We derived closed simple formulas for the maximum likelihood parameter estimation which constitutes a significant advantage of these distributions over negative binomial distribution.	
	
%Negative binomial distribution is the most used distribution to model macro-parasite burden in hosts. However reliable maximum likelihood parameter estimation from data is far from trivial. No closed formula is available and numerical estimation requires sophisticated methods. Using data from the literature we show that simple alternatives to negative binomial, like zero inflated geometric or hurdle distributions produce as good and even better fit to data  than negative binomial distribution.  We derived  closed simple formulas for the maximum likelihood parameter estimation which constitutes significant advantage of these distributions over negative binomial distribution. 


\end{abstract}

\tableofcontents

\section{Introduction}

%- acerca de macroparasitos y su importancia en ciencias bio-medicas.

Macroparasites usually present over-dispersed distributions where few hosts account for most of the parasites in the population (see, for example,  \citep{crofton1971quantitative}\citep{seo1979frequency}). 

The most used distribution is the negative binomial distribution (\citep{bliss91953}\citep{shaw1998patterns}) which provide and accurate description of the observations. 

However in many cases the negative binomial distribution (or other similar distributions) cannot account for the ``excess'' of zeros observed. A simple solution widely used is to consider zero-inflated distributions \citep{lambert1992zero}\citep{greene1994accounting}\citep{hall2000zero}.

The negative binomial distribution is a two parameters distribution, which usually 
are the mean burden of parasites in the host population $m$ and the inverse dispersion parameter $k$ which is related with the degree of the over-dispersion \citep{bliss91953}. Moreover, it can be shown that the limiting distribution of the $\mathrm{NB}(m,k)$ distribution, as $k\to \infty$, is a Poisson ($m$) distribution and if $k=1$ the $\mathrm{NB}(m,k)$ distribution is a geometric ($\frac{k}{m+k}$) distribution.

 %and when $k\to \infty$ becomes a Poisson ($m$) distribution.

However one problem with the negative binomial distribution is parameter estimation from the observations. The  method of moments estimation is simple but not always precise \citep{clark1989estimation}.  Maximum likelihood estimation (MLE) provides one of the best parameters estimation (\citep{piegorsch1990maximum}) but for the negative binomial distribution there is no a closed formula for the parameter estimates in terms of the observations and should be obtained numerically which present some complexities respect to the parameters numerical estimation for other distributions \citep{dai2013maximum}\citep{bandara2019computing}.

In this article we show alternatives to the negative binomial distribution which describe the observations equally well (and in some cases provide a most precise description) but for which we can compute a formula for the maximum likelihood parameters estimation values.

\section{Zero-inflated(deflated) distributions and hurdle distributions}

Macroparasites usually show over-dispersed distribution where few individuals have most of the parasite in the host population. It is said that distributions follow the 20-80 rule, 20\% percent of the individuals account for the 80\% of the parasite burden \citep{woolhouse1997heterogeneities}. Negative binomial distribution, among others,  offers such over-dispersed distribution but in some cases an excess of zeros is observed for which the distributions fail to account. A simple and widely used solution is to consider zero-inflated (in some cases zero-deflated) distributions and hurdle distributions \citep{WELSH1996297}\citep{doi:10.1191/1471082X05st084oa}.

\subsection{Zero-inflated(deflated) distributions}
 
A discrete random variable $Y$ follows a zero-inflated(deflated) distribution if its probability mass function $P$
is given by 	
\begin{equation}\label{zid}
P(Y=y)= \left\{ \begin{array}{lc}
\pi + (1-\pi)p(0;\theta) & y=0 \\
\\ (1-\pi)p(y;\theta)  & y\neq 0
\end{array}
\right.
\end{equation}
where we denote by $\theta$ to the vector of parameters of the associated distribution $p$ and then $$\sum_{y=0}^\infty p(y;\theta)=1.$$
When $\pi<0$ we have a zero-deflated distribution.

If $G(z)$ is the 
probability generating function
%genetating function 
of the distribution $p$, then the 
probability generating function
%generating function 
of the corresponding zero-inflated(deflated) distribution is given by
\begin{equation}
F(z)=\pi+(1-\pi)G(z)
\end{equation}
From the 
probability generating function
%moment generating function 
we may obtain the mean and the  variance for the distribution  straightforwardly
\begin{equation}\label{mzid}
\begin{split}
E(Y)&=(1-\pi)\mu\\
Var(Y)&=(1-\pi)\sigma^2+\pi(1-\pi)\mu^2
\end{split}
\end{equation}
where $\mu,$ $\sigma^2$  are the mean and variance of the distribution $p$. The coefficient of dispersion, 
or variance-to-mean ratio %(VMR) 
is therefore $D=\frac{Var(Y)}{E(Y)}$, 
 
\begin{equation}
D=\frac{\sigma^2}{ \mu}+\pi \mu
\end{equation}
where $\sigma^2/\mu$ is the variance-to-mean ratio %(VMR) 
for the associated distribution  $p$. As expected, zero-inflated distributions are more over-dispersed than the associated distribution.

%{\color{red}zero-inflated(deflated) distribution is } 

\subsection{Hurdle distributions.}

Another common way to account for an excess of zeros are the hurdle distributions, 
\begin{equation}\label{hd}
P(Y=y)= \left\{ \begin{array}{lc}
\pi & y=0 \\
\\ (1-\pi)\frac{p(y;\theta)}{1-p(0;\theta)}  & y\neq 0
\end{array}
\right.
\end{equation}

If $G(z)$ is the probability generating function for the associated distribution $p$, then the probability generating function for the hurdle distribution is
\begin{equation}
F(z)=\pi +(1-\pi)\frac{G(z)-p(0;\theta)}{1-p(0;\theta)}
\end{equation}
from where we obtain the mean and the variance of the hurdle distribution as 
\begin{equation}\label{mhd}
\begin{split}
E(Y)&=\alpha\mu\\
Var(Y)&=\alpha\sigma^2+\alpha(1-\alpha)\mu^2
\end{split}
\end{equation}
where $\alpha=\frac{1-\pi}{1-p(0;\theta)}$, and $\mu$, $\sigma^2 $  are the mean and variance of the associated distribution $p$. 
Finally the variance-to-mean ratio %(VMR) 
for the  hurdle distribution is given by 
\begin{equation}
D=\frac{\sigma^2}{ \mu}+(1-\alpha) \mu
\end{equation}  
%{\color{green}
If $\pi>p(0;\theta)$ the hurdle distribution are more over-dispersed than the associated distribution.

\section{Parameter estimation and Maximum likelihood}
A simple, but in  general inaccurate, way to fit the parameters of a distribution from a sample consist in the use of the sample moments. For example, the 
negative binomial distribution has two parameters which can be expressed in terms of the two first moments. While this method is quite simple, do not provide a reliable fit \citep{clark1989estimation} \citep{piegorsch1990maximum}. Usually, the method of choice is maximum likelihood \citep{piegorsch1990maximum}. However, maximum likelihood estimation (MLE) not always produce a closed formula and parameters need to be estimated numerically \citep{bliss91953}\citep{dai2013maximum}\citep{bandara2019computing}. In the following we pose the problem of maximum likelihood parameter estimation for zero-inflated and hurdle distributions. 
%The case of generalized, inflated and hurdle, distributions is presented in Appendix \ref{sec:appendix}. 

\subsection{Maximum likelihood estimation for zero-inflated(deflated) and hurdle distributions}

The problem of parameter estimation by maximum likelihood for zero-inflated and hurdle distributions is presented in the following. 
%The case of generalized inflated or hurdle distribution is presented in the Appendix \ref{sec:appendix}. 

\subsubsection{Zero-inflated(deflated) distribution}
%Ahora discutimos como estimar los parámetros of zero inflated/deflated model  for the Maximum Likelihood Estimation.
We denote by $\theta$ to the vector of parameters of the associated distribution $p$ and therefore the set of parameters for the zero-inflated(deflated) distribution is ($\pi,\theta$).
If $y_1,\ldots, y_N$ are $N$ observations, $N_0$ is the number of observations with zero counts, then the log-likelihood function is given by
%If $y_1,\ldots, y_N$ are $N$ observations and $N_0$ if the number of observations with zero counts, we denote by $\theta$ to the vector of parameters of the associated distribution $p$  and therefore the set of parameters for the zero-inflated/deflated distribution is ($\pi,\theta$). 
%Then the log-likelihood function is given by
\begin{equation}\label{mlezid}
\ell(\pi,\theta)= N_0 \ln \left[ \pi + (1-\pi)p_0\right] + (N-N_0) \ln (1-\pi) +\sum_{y_i\neq0} \ln p_{y_i}
\end{equation}
Maximizing $\ell$ for  ($\pi,\theta$) we obtain the following system, 
\begin{equation}\label{eqsmaxzid}
\begin{split}
\frac{\partial \ell}{\partial \pi}&=\frac{N_0 ( 1-p_0 ) }{\pi+(1-\pi)p_0}-\frac{N-N_0}{1-\pi}=0\\
\frac{\partial \ell}{\partial \theta_i}&=\frac{N_0(1-\pi)\dfrac{\partial p_0}{\partial \theta_i}}{\pi+(1-\pi)p_0}+\sum_{y_i\neq0} \frac{\frac{\partial  p_{y_i} }{\partial \theta_i}}{p_{y_i}}=0
\end{split}
\end{equation}
where we denote by $p_{y_i}=p(y_i;\theta)$.

\subsubsection{Hurdle distributions}
The  log-likelihood function for a hurdle distribution is given by
\begin{equation}\label{mlehd}
\ell(\pi,\theta)= N_0 \ln \pi + (N-N_0) \ln \left(\frac{1-\pi}{1-p_0} \right)  +\sum_{y_i\neq0} \ln p_{y_i}
\end{equation}
and therefore the parameter values are obtained from the system 
\begin{equation}\label{eqsmaxhd}
\begin{split}
\frac{\partial \ell}{\partial \pi}&=\frac{N_0}{\pi}-\frac{N-N_0}{1-\pi}=0\\
\frac{\partial \ell}{\partial \theta_i}&=\frac{(N-N_0) \dfrac{\partial p_0}{\partial \theta_i} }{1-p_0}+\sum_{y_i\neq0} \frac{\frac{\partial  p_{y_i} }{\partial \theta_i}}{p_{y_i}}=0
\end{split}
\end{equation}

%\section{Two simple distributions}

\section{Maximum likelihood for the negative binomial distribution and two simple alternatives.}

%{\color{red} aca mostrar el problema de la NBD, que ademas emperora si tenemos le caso inflado en cero. mostrar que no hay soluciones analiticas del problema y describir los problemas de la estimacion numerica, no puede usarse newton, los otros metodos son mas complicados y ademas pueden no encontarr el maximo... }

	The probability mass function for the negative binomial distribution is given by 
	\begin{equation}
	P(X=x)=\frac{\Gamma(x+k)}{\Gamma(x+1)\Gamma(k)}p^k q^x
	\end{equation}
	where $0\leq p\leq1$ y $k>0$. 
	Differentiating the log-likelihood function $\ell$ partially and setting
	them equal to zero yields the following system of likelihood equations  
	\begin{equation}
	\begin{split}
	\frac {\partial \ell }{\partial p}&=\frac {Nk}{p}-\frac {mN}{1-p}=0\\
	\frac {\partial \ell }{\partial k}&=\left[ \sum_i\psi (y_{i}+k)\right] -N\psi(k)+N\ln p=0
	\end{split}
	\end{equation}
	where $\psi (y)={\frac {\Gamma '(y)}{\Gamma (y)}}\!$ is the digamma function.
	Substituting in the second equation the value of $p=\frac{k}{m+k}$ obtained from the first equation, gives:
	\begin{equation}\label{nbmle}
	\left[ \sum_i\psi (y_{i}+k)\right] -N\psi(k)+N\ln \left(\frac{k}{m+k} \right) =0
	\end{equation} 
	This equation cannot be solved for $ k $ in a closed form and must be solved numerically. 
	Interative technique as Newton-Raphson method can be used, but this method may fail to find the MLE value.
	An analysis of the literature indicates that finding the MLE value is a challenge, since we could not obtain the root or obtain more than one for the equation (\ref{nbmle})
	\citep{clark1989estimation}\citep{piegorsch1990maximum}\citep{dai2013maximum}\citep{willson1984multistage}\citep{saha2005bias}.

\subsection{Zero-inflated geometric distribution}

The  geometric distribution is a special case of the negative binomial distribution for $k=1$, and then, its probability mass function is given by
\begin{equation}\label{geo}
P(X=x)= p q^x \quad \text{con $q=1-p$ } 
\end{equation}
where $x=0,1,2,\ldots$ and $0\leq p \leq1$.  The mean is $\frac{1-p}{p}$ while the variance is $\frac{1-p}{p^2}$.

\bigskip

From (\ref{zid}) the corresponding zero-inflated(deflated) distribution is 

\begin{equation}\label{zig}
P(Y=y)= \left\{ \begin{array}{lc}
\pi + (1-\pi)p & y=0 \\
\\ (1-\pi)p q ^y  & y\neq 0
\end{array}
\right.
\end{equation}

Mean and variance is given by  (\ref{mzid}), 
\begin{equation}
\begin{split}
E(Y)&=(1-\pi)\frac{(1-p)}{p}\\
Var(Y)&=(1-\pi)\frac{(1-p)}{p^2}[1+\pi(1-p)].
\end{split}
\end{equation}

The variance-to-mean ratio for the zero-inflated(deflated) geometric distribution is always greater than one, and therefore this distribution is always over-dispersed.

\subsubsection{Maximum likelihood estimation}
%{\color{red} aca borre algo....}
%The log likelihood function (\ref{mlezid}) for the zero inflated geometric distribution becomes
%\begin{equation}
%\ell(\pi,p)= N_0 \ln (\pi + (1-\pi)p)+ (N-N_0)\left[ \ln (1-\pi)+ \ln p \right]+ m N  \ln q
%\end{equation}
%where $m$  is the sample mean.  %{\color{red} aca falta algo... revisar}
System of likelihood equations is given according to (\ref{eqsmaxzid}) by 
\begin{equation}
\begin{split}\label{mlezig}
\frac{\partial \ell}{\partial \pi}&=\frac{N_0(1-p)}{\pi+(1-\pi)p}-\frac{N-N_0}{1-\pi}=0\\
\frac{\partial \ell}{\partial p}&=\frac{N_0(1-\pi)}{\pi+(1-\pi)p}+\frac{N-N_0}{p}-\frac{m N}{q}=0
\end{split}
\end{equation}
%where $m$  is the sample mean.
therefore, the best parameter estimations %obtained from (\ref{mlezig}) 
result
\begin{equation}
\begin{split}
\hat\pi&=\frac{mN_0-N+N_0}{mN-N+N_0}\\
\hat p&=\frac{N-N_0}{mN}
\end{split}
\end{equation} 
 which are in terms of the observed sample mean $m$, the sample size $N$ and the number of zeros in the sample $N_0$. 

\subsection{Hurdle geometric distribution}

The hurdle distribution for the associated geometric distribution is given by 
\begin{equation}\label{hgd}
P(Y=y)= \left\{ \begin{array}{lc}
\pi  & y=0 \\
\\ (1-\pi) \frac{pq ^y}{1-p}  & y\neq 0
\end{array}
\right.
\end{equation}

Mean and variance are obtained in a straightforward way, 
\begin{equation}
\begin{split}
E(Y)&=\frac{(1-\pi)}{p}\\
Var(Y)&=\frac{(1-\pi)}{p^2}\left[1+ (\pi-p)\right] 
\end{split}
\end{equation}

The variance-to-mean ratio for the hurdle geometric distribution is always greater than one, and therefore this distribution is always over-dispersed. 
%In this case the variance-to-mean ratio is given by $D=\frac{1+\pi-p}{p}$ and the distribution is overdispersed when  $\pi + 1 > 2p$.

\subsubsection{Maximum likelihood estimation}
%{\color{red} aca borre algo....}
%The log likelihood function (\ref{mlezid}) for the zero inflated geometric distribution becomes
%\begin{equation}
%\ell(\pi,p)= N_0 \ln (\pi + (1-\pi)p)+ (N-N_0)\left[ \ln (1-\pi)+ \ln p \right]+ m N  \ln q
%\end{equation}
%where $m$  is the sample mean.  %{\color{red} aca falta algo... revisar}
 According to (\ref{eqsmaxhd}) the system of likelihood equations is given by 
\begin{equation}
\begin{split}
\frac{\partial \ell}{\partial \pi}&=\frac{N_0}{\pi}-\frac{N-N_0}{1-\pi}=0\\
\frac{\partial \ell}{\partial p}&=(N-N_0)\left( \frac{1}{1-p}+\frac{1}{p} \right) - \frac{mN}{1-p}=0
\end{split}
\end{equation}
%where $m$  is the sample mean.
therefore, the best parameter estimations %obtained from (\ref{mlezig}) 
result
\begin{equation}
\begin{split}
\hat \pi&=\frac{N_0}{N}\\
\hat p&=\frac{N-N_0}{Nm}
\end{split}
\end{equation}
which are in terms of the observed sample mean $m$, the sample size $N$ and the number of zeros in the sample $N_0$.

%{\color{blue} 
	Note that the pair $\left( \dfrac{\hat\pi-\hat p}{1-\hat p},\hat p\right) $ is a maximum of log-likelihood function \eqref{mlezid}.	
	By the change of variables $\pi_{\mathrm{hg}}=\pi_{\mathrm{zig}} + (1-\pi_{\mathrm{zig}})p$
	the probability mass functions of the zero-inflated geometric (zig) and hurdle geometric (hg) models coincide
	%the adjustment
	%the fit by 
	%the zero-inflated geometric (zig) and hurdle geometric (hg) models will be the same 
	and the corresponding Akaike's information criterion values will be the same.
	Due to this, the fit of the data by both models coincide.
	%In what follows we will only carry out the analysis of the geometric model inflated to zero.
	In what follows we will only carry out the analysis of the zero-inflated geometric model.

%}

\section{Some examples}

\subsection{Study of frequency distribution of {\it Ascaris lumbricoides} infection}

The nematode {\it Ascaris lumbricoides} is one of the more commons intestinal parasites of humans. Highly prevalent in tropical and temperate populations where poverty and lack of sanitation is common \citep{pullan2012global}.

Burden of infestation is computed using the number of parasites in each host of the sample \citep{seo1979frequency}. At present time this type of studies are not longer conducted and we will use the results from Seo \citep{seo1979frequency}.  He studied six rural populations in Korea where an endemic situation was observed. 

\begin{figure}[h!]
	\centering
	\includegraphics[width=.99\linewidth]{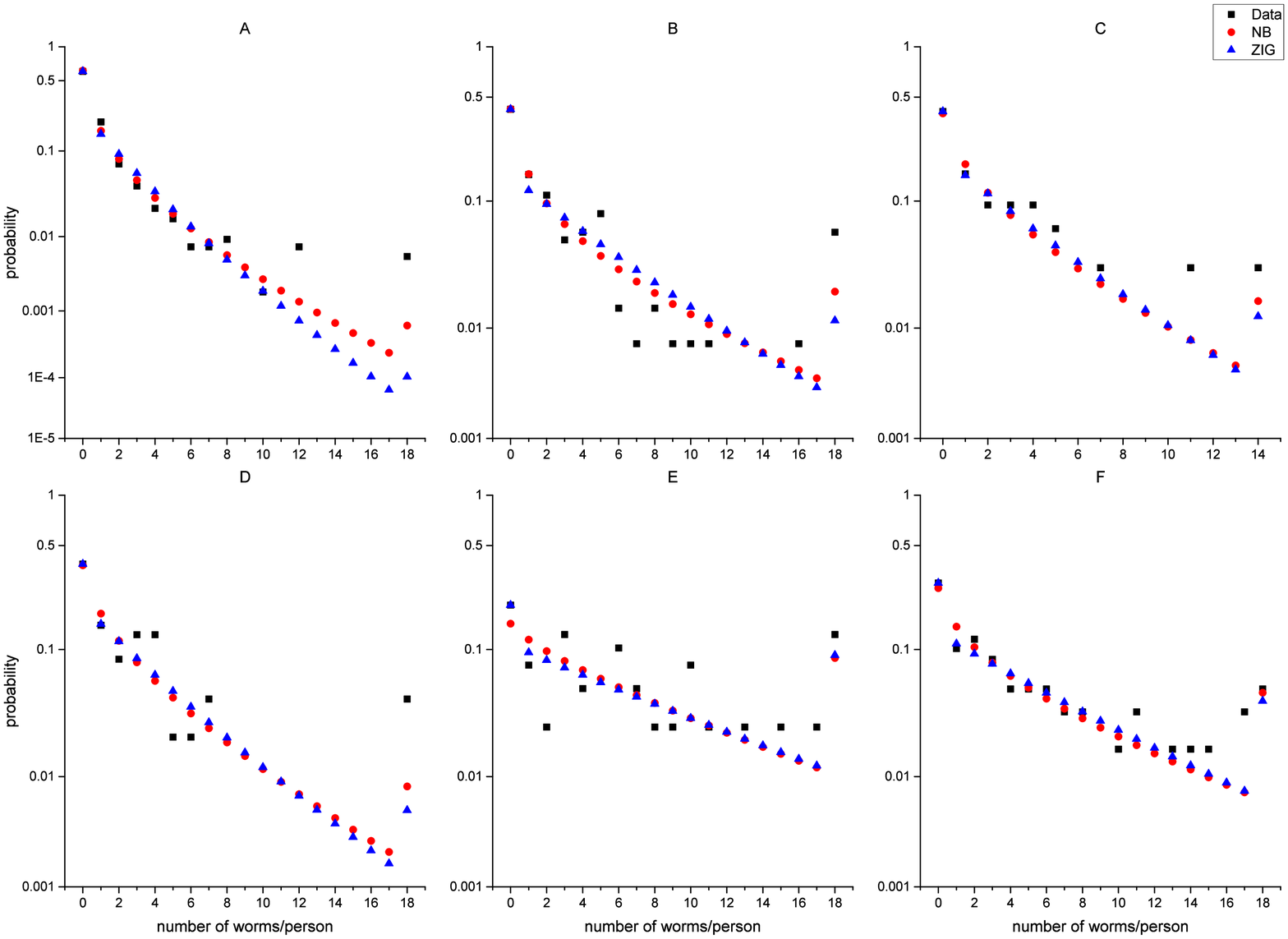}
	\caption{Fitting the parasite counts data (black) by NB (red) and ZIG (blue) distribution for Seo data set \citep{seo1979frequency}. Except in the first case, the simple zero-inflated geometric distribution fit the data as well as the negative binomial distribution (see Table \ref{table:seo})}
	\label{fig:seo}
\end{figure}

The samples showed over-dispersed distributions which could be accurately fitted by a negative binomial distribution (see Figure \ref{fig:seo}).
%\begin{figure}[h!]
%	\centering
%	\includegraphics[width=1.\linewidth]{seo.eps}
%	\caption{Fitting the parasite counts data (black) by NB (red) and ZIG (blue) distribution for Seo data set \citep{seo1979frequency}. Except in the first case, the simple zero inflated geometric distribution fit the data as well as the negative binomial distribution (see Table \ref{table:seo})}
%	\label{fig:seo}
%\end{figure}

%{\color{red}
In Figure \ref{fig:seo} we show the observed (black) and expected values of the fitted models (negative binomial and zero-inflated geometric). In addition, Table \ref{table:seo} includes the maximum likelihood estimations, the chi-squared statistics and their corresponding $p$-values, and Akaike's information criterion (AIC). 
As we see, the zero-inflated geometric distribution 
fit the data as well as the negative binomial distribution in most cases.
%seems to give a satisfactory fit in most cases. 
The AIC results show that the models negative binomial and zero-inflated geometric are similar.
Indeed, the fit in Figure \ref{fig:seo} improves the results obtained by negative binomial distribution in samples E and F. Using AIC, the model zero-inflated geometric showed the best performance in samples C, E and F.
Hence, the ZIG($\pi,p$) is a suitable candidate model to fit such data.
\subsection{Parasite distribution in crabs}
%\begin{figure}[h!]
%	\centering
%	\includegraphics[width=0.7\linewidth]{"Captura de pantalla de 2021-05-30 02-19-06"}
%	\caption{BORRAR}
%	\label{fig:captura-de-pantalla-de-2021-05-30-02-19-06}
%\end{figure}
Crofton \citep{crofton1971quantitative} contributed significanly to the study of parasite distributions in hosts. In his works he observed that over-dispersion is one of the main characteristics of parasite host distributions. Analyzing data from  Hynes and Nicholas \citep{hynes1963importance} of parasite infestation of the crustacean \textit{Gammarus pulex}, by the parasite %{\color{red} no me queda clara la s poblaciones... revisar paper original}
acanthocephalan \textit{Polymorphus minutus}.  Crofton show that negative binomial distribution provides a good fit to the data.

%{\color{red}
	
%In Figure \ref{fig:crofton}  we compare the fit to the data obtained by Crofton using the negative
%binomial distribution and  the zero-inated(deated) geometric distribution (Fig 2).	
	
\begin{figure}[h!]
	\centering
	\includegraphics[width=.99\linewidth]{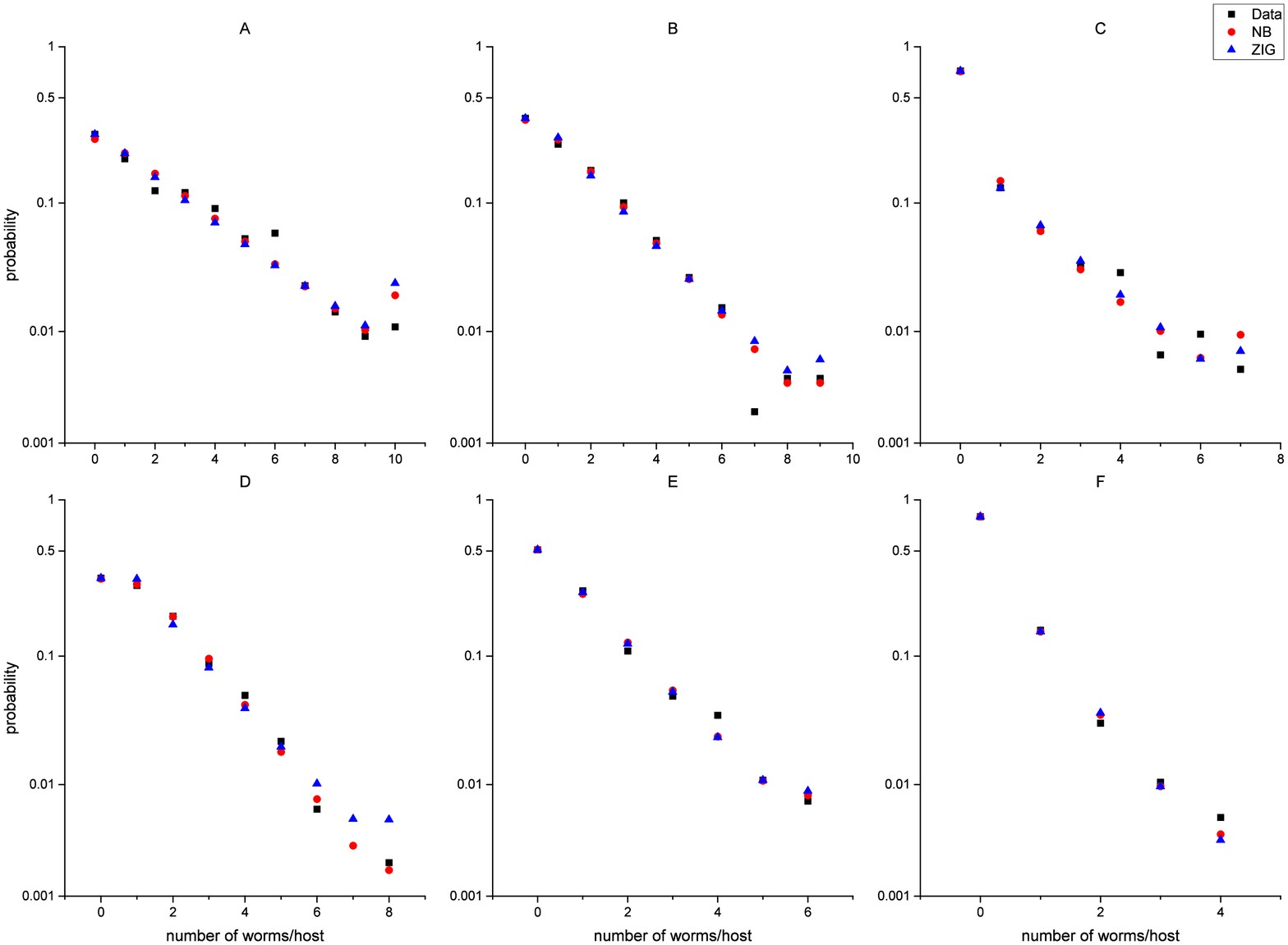}
	\caption{Fitting the parasite counts data (black) by NB (red) and ZIG (blue) distribution for  data in parasite in crabs \citep{crofton1971quantitative}. Zero-inflated(deflated) geometric distribution fit the data as well or better than the negative binomial distribution (see Table \ref{table:crofton})}
	\label{fig:crofton}
\end{figure}

In Figure \ref{fig:crofton} we compare the fit to the data obtained by Crofton using the negative
binomial distribution and  the zero-inflated(deflated) geometric distribution. %(Figure \ref{fig:crofton}).		
%In  Figure \ref{fig:crofton} we compare the results obtained by Crofton using the negative binomial distribution with the zero-inflated(deflated) geometric distribution. 
In most cases our simple proposal provides a better fit than the negative binomial distribution.%

%{\color{red} 
%The Fitting the observed and expected values is presente in Figure \ref{fig:crofton}. 
Based on the AIC and the chi-square goodness-of-fit test reported in Table \ref{table:crofton} 
%it is obvious that the proposed 
we conclude that the zero-inflated(deflated) geometric distribution provides a fit of the data as good as the negative binomial distribution.
%ZIG($\pi,p$), gives good fit to the data sets considered. 
%}

%\begin{figure}[h!]
%	\centering
%	\includegraphics[width=1\linewidth]{crofton1-3}
%	\caption{}
%	\label{fig:crofton1-3}
%\end{figure}
%\begin{figure}[h!]
%	\centering
%	\includegraphics[width=1.\linewidth]{crofton.eps}
%	\caption{Fitting the parasite counts data (black) by NB (red) and ZIG (blue) distribution for  data in parasite in crabs \citep{crofton1971quantitative}. Zero inflated geometric distribution fit the data as well or better than the negative binomial distribution (see Table \ref{table:crofton})}
%	\label{fig:crofton}
%\end{figure}

%\newpage
\section{Discussion and Conclusions}

%{\color{red} 
Negative binomial distribution is widely used to describe parasite burden in populations. It provides good fits of the observations which can be improved by the corresponding zero-inflated(deflated) distribution \citep{crofton1971quantitative} \citep{seo1979frequency}.

However parameters estimation is far from trivial and maximum likelihood estimates must be found always numerically. %{\color{green} 
Simple numerical methods (as the Newton method) are not easy to implement as there is not a closed expression for the derivative of the Gamma function and fail if the starting value is not chosen appropriately \citep{bandara2019computing}.
		
%}

	Negative binomial distribution may also fail to fit the zero counts \citep{crofton1971quantitative}. This may problem may be overcome using the zero-inflated negative binomial distribution but in this case parameters estimation by maximum likelihood is even more complex 
	and also the AIC criteria penalize models with a larger number of parameters.

	For zero-inflated geometric distribution or hurdle geometric distribution we found simple formulas for the maximum likelihood parameter estimates.

	In the examples analyzed in this work zero-inflated geometric distribution or hurdle geometric distributions present in most cases a similar fit to the data than the negative binomial distribution. 
	In the few cases this distribution significantly improve the fit. The AIC results show that the models are similar.

	%In the few cases where the fit is not good enough one-count inflation/deflation or the corresponding hurdle distribution significantly improve the fit. 
	However the major advantage of this distributions is not a little improvement in the data fitting but the fact that simple formula is provided for the  maximum likelihood parameter estimates. 
	
%}

A simple formula for the distribution's parameters, avoiding the use of complex numerical methods for parameter estimation, may result of practical convenience for many researchers working in the area which are not familiar with programming and the  numerical implementation of algorithms.

For all the models considered not good fit of the tail of the distribution is observed. However this fact does not necessarily indicates the need to consider other distributions. Because samples are small, rare events in the tail, when observed acts like outliers given too much weight to this rare observation.

\section*{Aknowledgements}

This work was partially supported by grant CIUNSA 2018-2467. JPA is a member of the CONICET. GML is a doctoral fellow of CONICET.

\section*{Conflict of Interest}

The authors have declared no conflict of interest.

%\newpage

%\bibliographystyle{ama}
%\refname{Reference}
%\bibliographystyle{apa}
%\renewcommand{\refname}{References}  % for the article class
%\printbibliography[title={Reference}]
\bibliographystyle{apa}
\bibliography{biblio}	

%\newpage
\section{Tables}
%tabla 1 corregida
\begin{table}[h!]
	\centering
	\resizebox{\textwidth}{!}{%
		\begin{tabular}{cccccccc}
			\\
			\hline 
			\multirow{3}{*}{\minitab [c]{Theoretical\\distribution}} & \multirow{3}{*}{\minitab [c]{Calculated\\parameters}} & \multicolumn{6}{c}{Samples} \\ \cline { 3-8 }
			&                            & A  & B & C  & D  & E  & F \\
			&                            & ($n=540$)  & ($n=136$)  & ($n=32$)  & ($n=47$)  & ($n=39$)  & ($n=59$) \\
			\hline
			\multirow{4}{*}{NB}  & $m$     &  1.0167 &  2.8235  &  2.3125  &  2.5106  &  6.6410  &  4.6102  \\
			& $k$                          & 0.3546 & 0.4240 & 0.5761 & 0.5893 & 0.8726 & 0.6193 \\
			& chi-squared statistic        & 50.0660 & 25.8883 & 6.8092 & 17.4885 & 15.6914 & 11.2927 \\
			%& df                         &  16   &    &    &    &    &   \\
			& $p$-value                     &  $<0.0001$  &  0.0556  &  0.8699  &   0.3547  &  0.4747  &  0.7911 \\
			& AIC                     & \textbf{ 1438.0235}  &  \textbf{574.5621}  &  131.7761  &   \textbf{198.1410}  &  235.0388  &  310.0239  \\

			\hline
			\multirow{4}{*}{ZIG}   & $\pi$ &  0.3653   &  0.2687  &  0.2011  &  0.1819  &  0.0971  &  0.1581 \\
			& $p$                          &  0.3843  &  0.2057  & 0.2568  &  0.2458  &  0.1197  &  0.1544 \\
			& chi-squared statistic        &  213.1743   &  43.2591  &  6.8197  &  22.2336  &  14.4552  &  10.6473 \\
			%& df                         &  16   &    &    &    &    &   \\
			& $p$-value                     &  $<0.0001$   &  0.0003  &  0.8693  &  0.1358  &  0.5648  & 0.8307 \\ 
			& AIC                     &  1458.0335   &  579.2921  &  \textbf{131.5360}  &  198.1585  &  \textbf{233.3277}  &  \textbf{308.9315}  \\

			%			\hline
			%			\multirow{4}{*}{ZIP}   & $\pi$ &  0.3653   &  0.2687  &  0.2011  &  0.1819  &  0.0971  &  0.1581 \\
			%			& $\lambda$                          &  0.3843  &  0.2057  & 0.2568  &  0.2458  &  0.1197  &  0.1544 \\
			%			& chi-squared statistic        &  213.1743   &  43.2591  &  6.8197  &  22.2336  &  14.4552  &  10.6473 \\
			%			%& df                         &  16   &    &    &    &    &   \\
			%			& $p$-value                     &  $<0.0001$   &  0.0003  &  0.8693  &  0.1358  &  0.5648  & 0.8307 \\
			%			& AIC                     &  1458.0335   &  0.0003  &  0.8693  &  0.1358  &  0.5648  & 0.8307 \\
			
			\hline 
			& df                     &  16   &  16  &  12  &  16  &  16  & 16
		\end{tabular} 
	}%
	\caption{Parameters of NB and ZIG distributions calculated from observed Seo data \citep{seo1979frequency}
		%(\citep{seo1979frequency}) 
		and results of chi-squared test and AIC}
	\label{table:seo}
\end{table}

%tabla2corregida
\begin{table}[h!]
	\centering
	\resizebox{\textwidth}{!}{%
		\begin{tabular}{cccccccc}
			\\
			\hline 
			\multirow{3}{*}{\minitab [c]{Theoretical \\ distribution}} & \multirow{3}{*}{\minitab [c]{Calculated  \\ parameters}} & \multicolumn{6}{c}{Samples}  \\ \cline { 3-8 }
			&                            & A  & B & C  & D  & E  & F \\
			&                            & ($n=549$)  & ($n=509$)  & ($n=633$)  & ($n=486$)  & ($n=276$)  & ($n=191$) \\
			\hline
			\multirow{4}{*}{NB}           & $m$          &  2.2732   &  1.4165  &  0.6003  &  1.3189  &  0.8913  &  0.2670 \\
			& $k$                          & 1.2564 & 1.5837 & 0.2974 & 3.0544 & 1.2679 & 0.6069 \\
			& chi-squared statistic        & 20.6558 &  3.1086 &  10.5075  &  2.9993 &  2.3843  &  0.2776 \\
			%& df                         &  16   &    &    &    &    &   \\
			& $p$-value                     &  0.0081  &  0.8748  &  0.0621  &   0.8089  &  0.6655  &  0.8704 \\
			& AIC                     &  \textbf{2211.4460}  &  \textbf{1662.1623}  &  1279.5742  &  \textbf{1506.5751}  &  724.9286  &  \textbf{252.4139} \\
			\hline
			\multirow{4}{*}{ZIG}          & $\pi$      &   -0.0256  &  -0.1304  &  0.4875  &  -0.3313  &  -0.1020  &  0.2195 \\
			& $p$                          &  0.3109   &  0.4438  &  0.4605  &  0.5023  &  0.5528  &  0.7451 \\
			& chi-squared statistic                        &  23.4185   &  6.0467  &  6.5825  &  8.6706  &  2.1793  &  0.4542 \\
			%& df                         &  16   &    &    &    &    &   \\
			& $p$-value                     &  $0.0029$   &  $0.5343$  &  0.2536  &  0.1930  &  0.7028  & 0.7969 \\
			& AIC                     &  2215.4026   &  1665.9254  &  \textbf{1274.8293}  &  1514.0280  &  \textbf{724.8346}  & 252.4989 \\
			\hline 
			& df                    &  8   &  7  &  5  &  6  &  4  & 2
		\end{tabular} 
	}%
	\caption{Parameters of NB and ZIG distributions calculated from observed of parasite in crabs data \citep{crofton1971quantitative} %(\citep{hynes1963importance}) 
		and results of chi-squared test and AIC.}
	\label{table:crofton}
\end{table}

\end{document}